# Low-voltage Ferroelectric Field-Effect Transistors with Ultrathin Aluminum Scandium Nitride and 2D channels

Chloe Leblanc[1], Hyunmin Cho[1], Yinuo Zhang[1], Seunguk Song[1,2,3], Zachary Anderson[1], Yunfei He[1], Chen Chen[4], Joan M. Redwing[4], Roy H. Olsson, III[1], Deep Jariwala[1, *]

[1]Department of Electrical and Systems Engineering, University of Pennsylvania, Philadelphia, Pennsylvania, USA
[2]Department of Energy Science, Sungkyunkwan University (SKKU), Suwon, 16419, Republic of Korea
[3]Center for 2D Quantum Heterostructures (2DQH), Institute for Basic Science (IBS), Sungkyunkwan University (SKKU), Suwon, 16419, Republic of Korea
[4]2D Crystal Consortium–Materials Innovation Platform, Materials Research Institute, Pennsylvania State University, University Park, Pennsylvania, USA

[*]Author to whom correspondence should be addressed: dmj@seas.upenn.edu.

**Abstract**

The continued evolution of CMOS technology demands materials and architectures that emphasize low-power consumption, particularly for computations involving large-scale data processing and multivariable optimization. Ferroelectric materials offer promising solutions through enabling dual-purpose memory units capable of performing both storage and logic operations. In this study, we demonstrate ferroelectric field-effect transistors (FeFETs) with MoS$_2$ monolayer channels fabricated on ultrathin 5 nm and 10 nm ferroelectric Aluminum Scandium Nitride (Al$_{1-x}$Sc$_x$N) films. By decreasing the thickness of the ferroelectric film, we achieve significantly reduced gate voltages (<3V) required to switch the conductance of the devices, enabling operation at low voltages compatible with advanced CMOS. We observe a characteristic crossover in hysteresis behavior that varies with film thickness, channel fabrication method, and environmental conditions. Through systematic investigation of multiple parameters including channel fabrication methods, dimensional scaling, and environmental effects, we provide pathways to improve device performance. While our devices demonstrate clear ferroelectric switching behavior, further optimization is required to enhance the ON/OFF ratio at zero gate voltage while continuing to reduce the coercive field of these ultrathin films.

**Introduction**

Ferroelectric materials exhibit the fundamental property of maintaining a stable polar state in the absence of an external electric field, making them attractive candidates for next-generation low-power memory and logic applications. The polarization acts as a memory of an electric field the material was previously exposed to and modulates the conductance of the film accordingly. When incorporated into a logic unit, the ferroelectric material in turn maintains the state of the device without the need for continuous application of an electric field. The unit is therefore less power-consuming. Establishing a specific polarization state requires aligning the material's dipoles through the application of a characteristic electric field, known as the coercive field ($E_C$), resulting in a uniform internal field[1,2]. This coercive field can be modulated through various mechanisms, including doping and dimensional scaling. For instance, piezoelectric aluminum nitride (AlN) transitions to ferroelectric behavior when substitutionally alloyed with a minimum of 14% scandium, forming aluminum scandium nitride ($Al_{1-x}Sc_xN$)[3,4,5,6]. Since the discovery of ferroelectricity in $Al_{1-x}Sc_xN$, this material has attracted much attention for its exceptional remnant polarization and low growth temperatures, which are crucial for practical foundry-level applications. It also shows relatively high $E_C$ ∼3–5 MV/cm, which indicates higher operating voltages[7,8]. The thickness of ferroelectric films significantly influences their operating voltages, with thinner films typically exhibiting lower coercive fields[9,10,11]. For this reason, achieving thin $Al_{1-x}Sc_xN$ based devices have been a point of interest in recent studies[12,13,14,15].

In this work, we demonstrate field-effect transistors utilizing 10 nm thick $Al_{0.68}Sc_{0.32}N$ and 5 nm thick $Al_{0.72}Sc_{0.28}N$ films that achieve low switching voltages. Ferroelectric field-effect transistors (FeFETs) are a non-volatile architecture where the dielectric layer is substituted for a ferroelectric material that controls the channel conductance[8]. We employ two-dimensional (2D) molybdenum disulfide ($MoS_2$) as the semiconducting channel material, leveraging the demonstrated advantages of atomically thin 2D van der Waals semiconductors including superior lateral scaling capabilities, low in-plane dielectric constants, high ON-current density, and high-speed operation characteristics. By systematically exploring multiple device parameters including channel fabrication methods, dimensional scaling effects, and environmental conditions, we provide insights into optimization strategies for these promising ferroelectric devices. Our results demonstrate that ultrathin $Al_{1-x}Sc_xN$ can enable low-voltage ferroelectric switching, which is essential for future energy-efficient computing paradigms[16,17,18,19,20].

**Results**

The ferroelectric layer was sputtered onto a 50 nm (111)-oriented Al layer, which served as the back gate electrode. The complete device structure comprised an $Al/Al_{1-x}Sc_xN/2D\ MoS_2$ stack fabricated on sapphire substrates (Fig. 1(a)), with global back gates and source and drain contacts consisting of 10/40 nm In/Au metallization. This choice was particularly important as short channel 2D FETs are commonly limited by contact resistance[21,22]. Indium has a small work function of ∼4.1 eV, enabling an ohmic contact to create n-type FETs and minimal hole conduction at below $V_{GS}$= -2 V[7]. A signature behavior of FeFETs during bidirectional scanning transfer measurements is the gradual increase(decrease) of the gate voltage leads to a competition between ferroelectric polarization switching and depolarization caused by an imbalance between the applied and internal fields. When the gate voltage $V_{GS}$=0 V, the internal field is still oriented downwards(upwards)[2]. This behavior manifests as direction-dependent drain current characteristics: the current exhibits lower values during negative-to-positive gate voltage sweeps compared to positive-to-negative sweeps, producing counterclockwise hysteresis phenomenon[23]. Our devices demonstrate this counterclockwise behavior partially, with a consistent observation of a crossover point where the hysteresis transitions to clockwise behavior. This crossover is observed regardless of gate voltage range ($V_{GS}$ = ± 0.5 V, ± 1 V, ± 1.5 V, ± 2 V, ± 3 V, ± 4 V) or applied drain current ($V_{DS}$ = 0.1 V, 0.5 V, 1 V), and indicates a systematic shift in the turn-on voltage toward positive values (Fig. 1). The observed channel breakdown voltages are typically at $V_{GS}$ ≥ 5V due to joule heating from carrier injection. This prevents going to larger gate switching ranges, where the ferroelectric effect should be more dominant[7].

We verify that this behavior is still caused by ferroelectric polarization by recreating our devices on undoped 10 nm AlN layers (Fig. 2(d)). Since AlN film is non-ferroelectric (it would breakdown before switching under our measurement conditions), the devices display a purely clockwise hysteresis behavior. The clockwise pattern can be attributed to charge trapping phenomena at the dielectric/MoS$_2$ and metal/MoS$_2$ interfaces, resulting in the formation of interfacial dead layers and subsequent destabilization of the ferroelectric polarization[24].

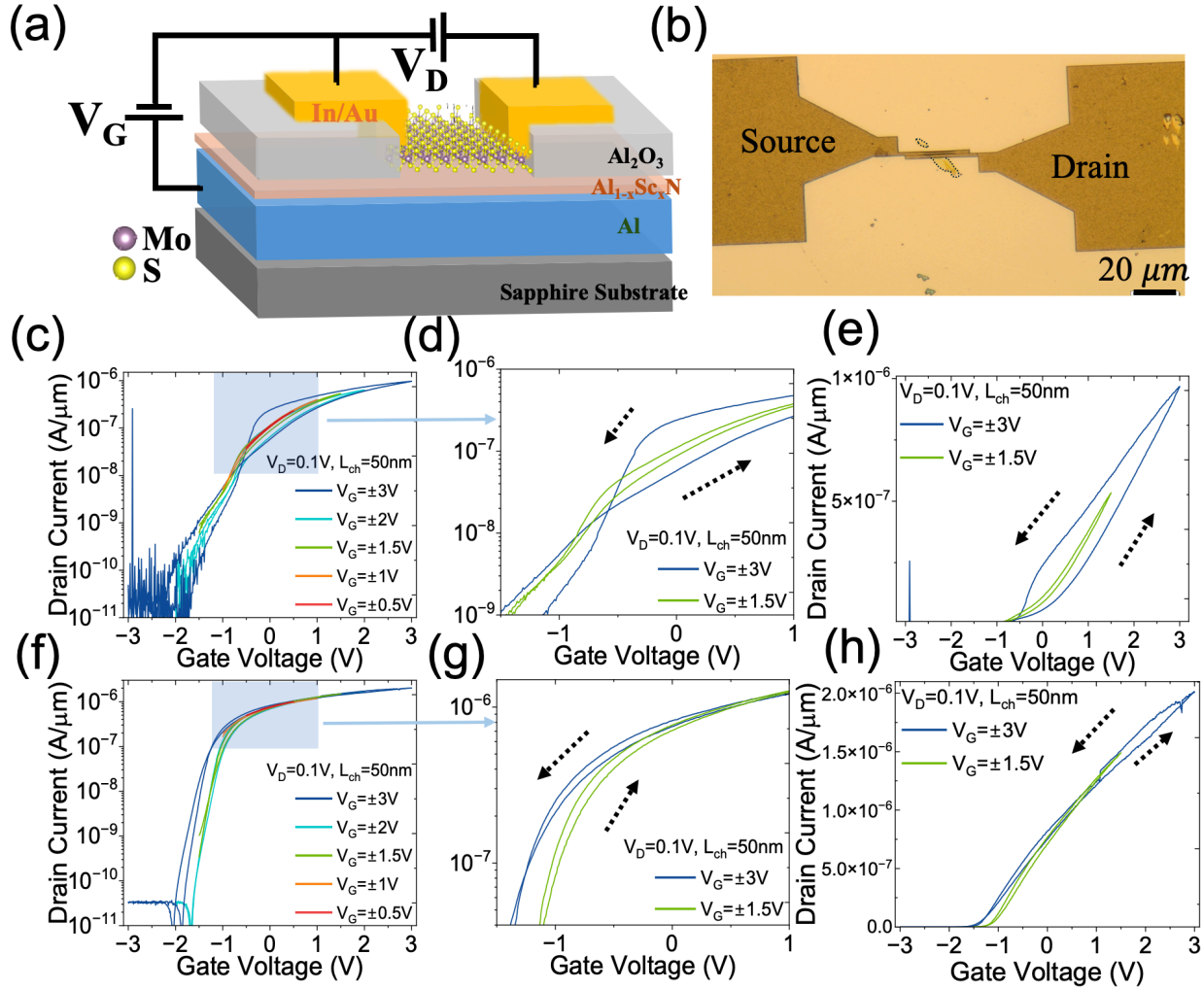

**Figure 1** (a) Schematic of the device structure (b) Optical microscope image of a 50 nm channel ($L_{ch}$) device on 5 nm Al$_{0.72}$Sc$_{0.28}$N. The dry-transferred semiconductor channel flake (~3 nm thick) is outlined with black dots. Active area is surrounded by a uniform insulating Al$_2$O$_3$ layer (see Methods section). 50 nm $L_{ch}$ flake MoS$_2$ device on (c-e) 10 nm thick Al$_{0.68}$Sc$_{0.32}$N (f-h) 5 nm thick Al$_{0.72}$Sc$_{0.28}$N. (c,f) Overlapping transfer curves of different $V_{GS}$ ranges normalized by channel area on logarithmic scale. $I_{DS}$-$V_{GS}$ and $I_{GS}$-$V_{GS}$ characteristics under applied $V_{DS}$ =0.1V, 0.5V, 1V and $V_{GS}$ = ±0.5V through ±3V are plotted individually in Fig. S1 and S2. (d,g) Zoom-in view of the $V_{GS}$ = ± 3V and $V_{GS}$ = ±1.5V ranges in the blue box of Fig. 1(c,f). Dotted arrows represent the sweep direction of $I_{DS}$. (e,h) Overlapping transfer curves of $V_{GS}$ = ±3V and $V_{GS}$ = ±1.5V ranges on linear scale. Output characteristics are available in Fig. S3.

Rapid charge trapping and releasing occurs in gate voltage ranges with high rectification ratio. In our 5 nm thick $Al_{0.72}Sc_{0.28}N$ exfoliated FeFET with 50 nm channel length in Fig. 2(a), the ideality factor averages to 4.07 and 3.85 during forward and backward sweeps in the range $V_{GS}$ = [-1.95; -1.40] V (see Fig. 2(a), S4). We also see an $I_{ON}/I_{OFF}$ ~$10^5$ within a relatively narrow $V_{GS}$ sweep range of ± 3V. Assuming the ferroelectric turn-on $V_{ON}$ voltage to be the crossover voltage, the transconductance of this device peaks at $V_{GS}$ - $V_{ON}$ = -0.32 V (see Fig. S5). Due to the difference in thickness, we also expect the sheet carrier density in the 5 nm thick $Al_{0.72}Sc_{0.28}N$ devices to be lower than in the 10 nm thick $Al_{0.68}Sc_{0.32}N$, which translates to a higher effective mobility[25]. This justifies the higher ON current values we observe irrespective of channel length in MOCVD wet-transferred $MoS_2$ devices (see Fig. S6). In addition, the counterclockwise hysteresis region becomes dominant as the $Al_{1-x}Sc_xN$ thickness is reduced. We see that for 50 nm channel devices on 10 nm thick $Al_{0.68}Sc_{0.32}N$ FeFETs, the crossover occurs at approximately gate voltage $V_{GS}$ = 0 V for an applied drain voltage $V_{DS}$ = 1 V and $V_{GS}$ = ±2V, whereas this decreases to $V_{GS}$ < -1.2 V for 5 nm thick $Al_{0.72}Sc_{0.28}N$ FeFETs under the same conditions (see Fig. S7). We verify that this behavior is not attributable to leakage by consistently recording the gate current $I_{GS}$ during measurements (see Fig. S1-S2, S8-S14).

However, the 5 nm thick $Al_{0.72}Sc_{0.28}N$ FeFET's main limitation is a reduced hysteresis window[2]. This limitation could stem from a native surface oxidation of the $Al_{1-x}Sc_xN$ layer forming during ambient air exposure, yielding $(AlSc)_xO_y$ at the surface[26,14]. To minimize this effect, the FeFETs are stored an argon environment with oxygen levels kept below 20 ppm prior to and during transfer (see Methods section)[27]. The oxidation increases the gate capacitance, which is inversely related to the memory window[28]. In contrast, we see a larger memory window in 10 nm thick $Al_{0.68}Sc_{0.32}N$, in which an oxidation layer is proportionally less significant. In contrast, their memory window is much larger (see Fig. 2(b,c)). For a constant applied drain voltage $V_{DS}$ = 0.1 V and $V_{GS}$ = ±3V, the ON/OFF ratio at $V_{GS}$ = 0V increases from 1.1 V to 4.13.

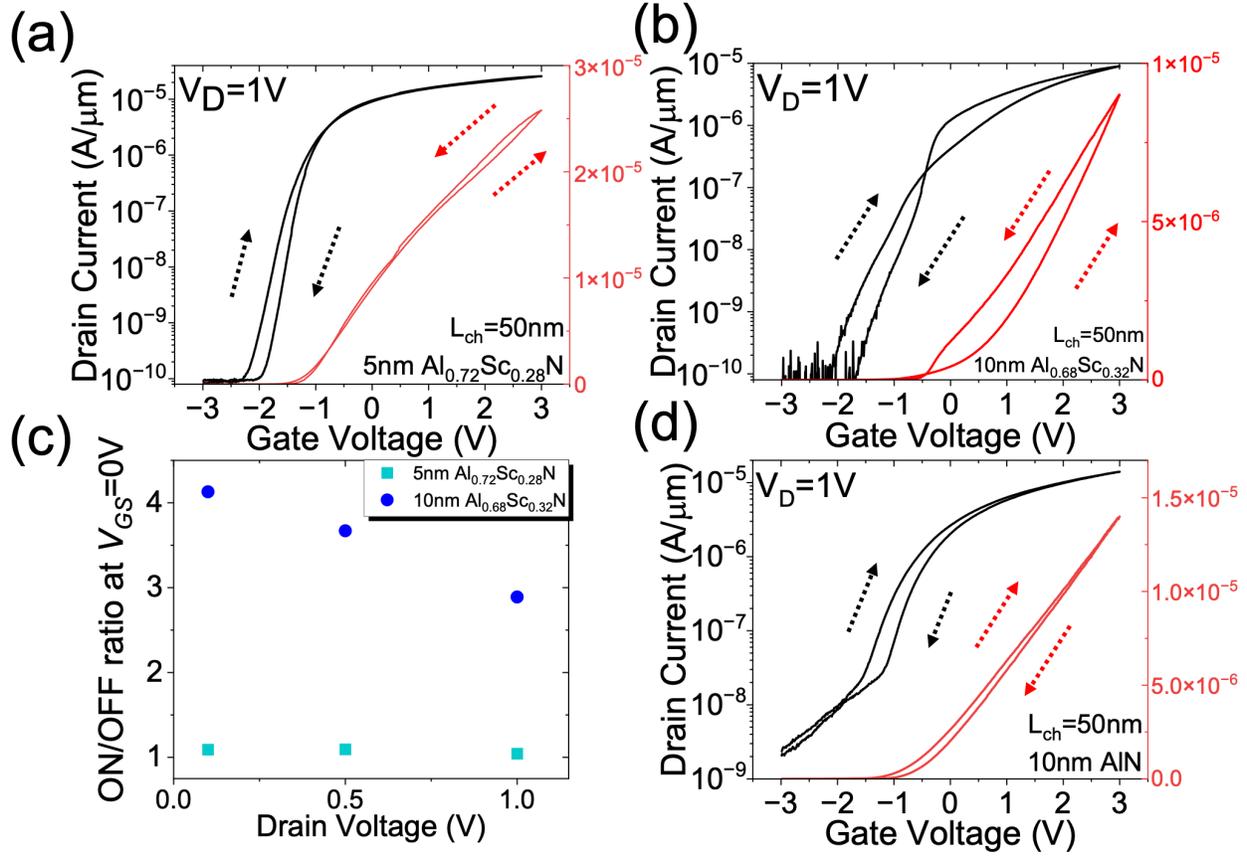

**Figure 2** Transfer characteristics of flake MoS$_2$, 50 nm channel (L$_{ch}$) devices with an applied $V_{DS}$ =1V on (a) 5 nm Al$_{0.72}$Sc$_{0.28}$N (b) 10 nm Al$_{0.68}$Sc$_{0.32}$N. Curves are plotted in logarithmic scale on the left axis (black) and linear scale on the right axis (red). Dotted arrows represent the sweep direction of $I_{DS}$. $I_{GS}$-$V_{GS}$ and $I_{DS}$-$V_{GS}$ characteristics under applied $V_{DS}$ =0.1V, 0.5V, 1V and $V_{GS}$ = ±0.5V through ±3V are available in Fig. S1 and S2. (c) ON/OFF ratio at $V_{GS}$ =0V of each device under applied $V_{DS}$ = 0.1V, 0.5V, 1V. (d) Transfer characteristics of a dry-transferred 50 nm channel (L$_{ch}$) device with an applied $V_{DS}$ = 1V on 10 nm AlN. Dotted arrows represent the sweep direction of $I_{DS}$. The higher OFF current comparatively to Fig. 1(b) is attributed to flake-to-flake thickness variations.

The effect of the MoS$_2$ transfer method onto the Al$_{1-x}$Sc$_x$N surface is also crucial to device performance. To study this, we compare FeFETs made via wet and dry transfer processes using MOCVD grown monolayer MoS$_2$. The chemical wet-transfer approach, outlined in the Methods section, introduces more impurities such as potassium ions at the interface between the channel and the ferroelectric dielectric than the dry-transfer technique, and these consistently yield a higher crossover voltage. In Fig. 3, devices with wet- and dry-transferred MOCVD MoS$_2$ on 10 nm Al$_{0.68}$Sc$_{0.32}$N are shown and compared with the flake MoS$_2$ on 10 nm Al$_{0.68}$Sc$_{0.32}$N device shown in Fig. 1, 2. In addition to having the lowest crossover voltages, the dry-transferred monolayer MOCVD MoS$_2$ channels present the highest ON/OFF ratios. For an applied $V_{DS}$ = 1V and channel length 500 nm, the ON/OFF ratio at $V_{GS}$ = 0V is 4.14 and the maximum memory window is approximately 0.064 V/nm at 2.6 x 10$^{-7}$ A/um, comparable to previously reported 2D-channel/ferroelectric FeFETs (~0.01–0.14 V/nm) [29,21,7]. The difference caused by transfer methods is also independent of channel length or drain voltage (see Fig. S1-S2, S8-S14). Given that shorter channels yield devices that are more prone to the effects of contacts, this supports the conjecture that the presence of the crossover voltage is related to the trapping of charges, which are notably caused by impurities distributed across the channel[30,22].

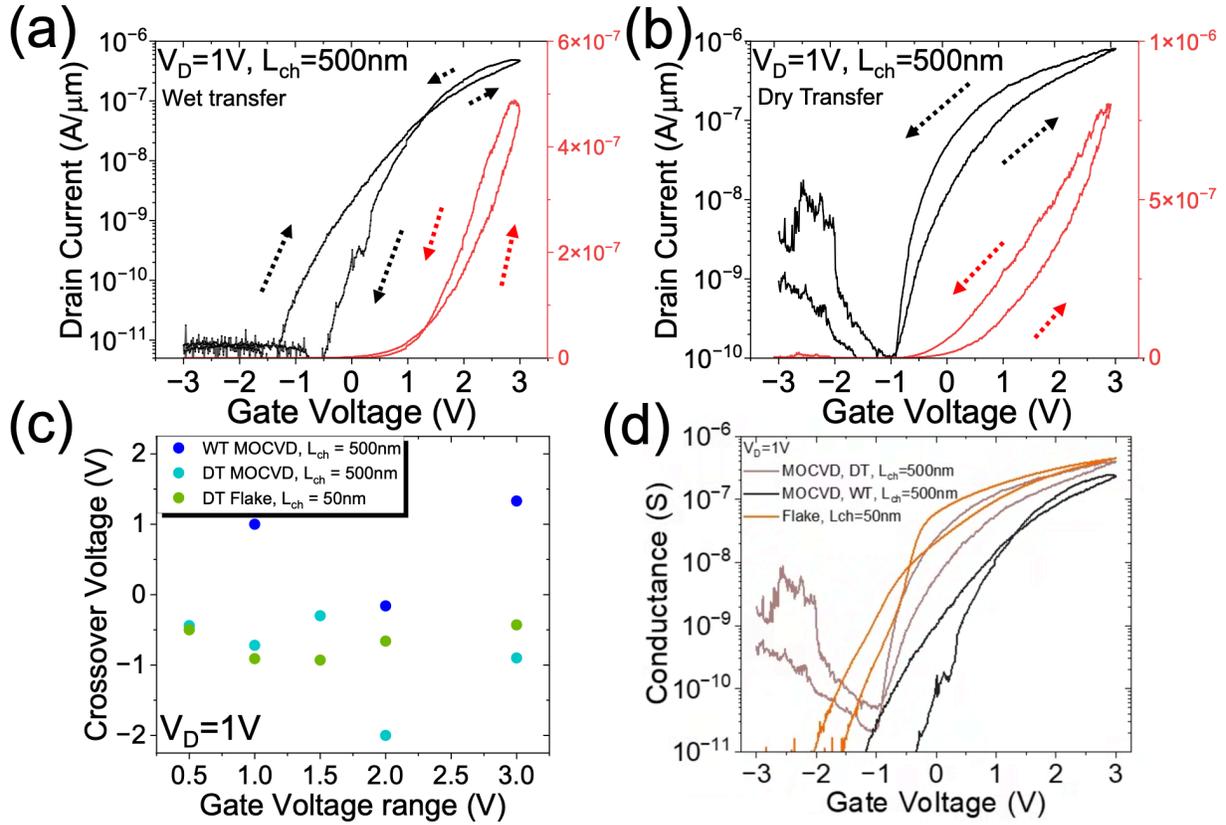

**Figure 3** Transfer characteristics of (a) wet- (b) dry-transferred 500 nm channel ($L_{ch}$) devices with MOCVD MoS$_2$ channels and an applied $V_{DS}$ =1V on 10 nm Al$_{0.68}$Sc$_{0.32}$N. Curves are plotted in logarithmic scale on the left axis (black) and linear scale on the right axis (red). Dotted arrows represent the sweep direction of $I_{DS}$. $I_{GS}$-$V_{GS}$ and $I_{DS}$-$V_{GS}$ characteristics at $V_{GS}$ =±0.5V through ±3V are available in Fig. S8. (c) Crossover voltage values at $V_{DS}$ = 1V of both devices under different $V_{GS}$ ranges. Crossover voltage values of the flake MoS$_2$ device presented in Fig. 1(c-e) is added for comparison. (d) Sheet conductance characteristics of the devices in Fig. 4(a-c). $I_{GS}$-$V_{GS}$ and $I_{DS}$-$V_{GS}$ characteristics of a wet-transferred 50 nm channel ($L_{ch}$) devices with MOCVD MoS$_2$ under applied $V_{DS}$ = 0.1V, 0.5V, 1V and $V_{GS}$ = ±1V through ±4V is available in Fig. S9 for comparison.

In addition, we verify that the improved crossover voltage occurs regardless of Al$_{1-x}$Sc$_x$N thickness. In Fig. 4, we compare devices on 5 nm Al$_{0.72}$Sc$_{0.28}$N with few-layer flake and MOCVD wet-transferred MoS$_2$ at channel lengths 100 nm, 250 nm and 500 nm for various drain voltages. Crossover voltages for flake channel devices consistently yield larger counterclockwise regions.

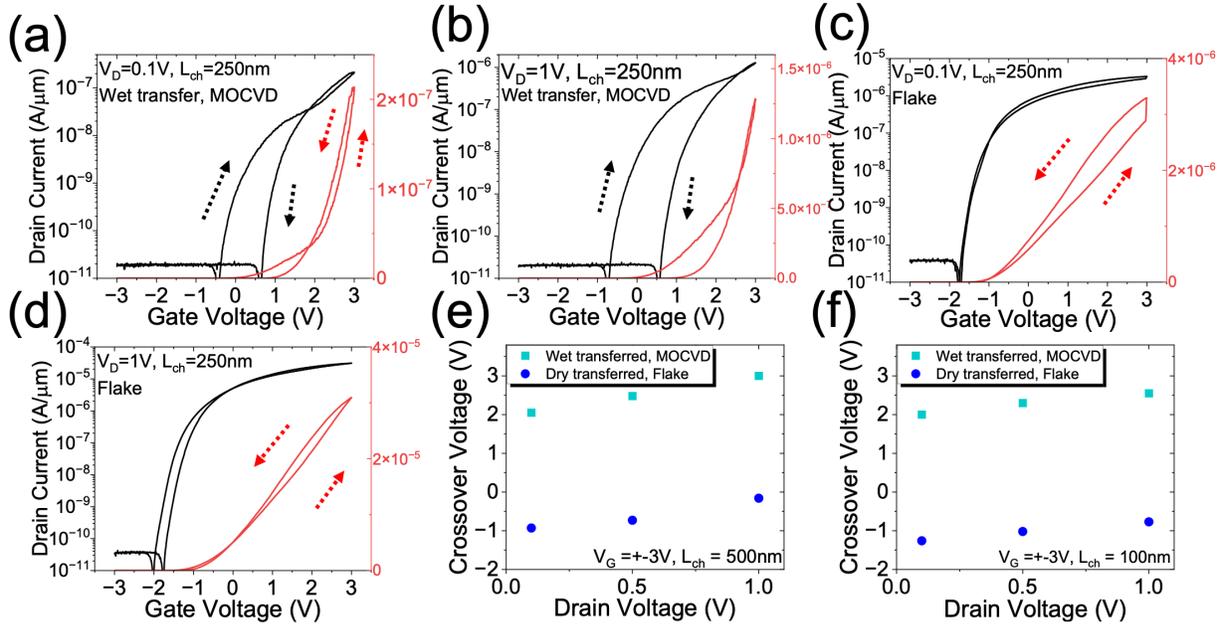

**Figure 4** Transfer curves of MOCVD MoS$_2$ wet-transferred 250 nm channel (L$_{ch}$) devices on 5 nm Al$_{0.72}$Sc$_{0.28}$N with an applied (a) $V_{DS}$ = 0.1V (b) $V_{DS}$ = 1V. Dotted arrows represent the sweep direction of $I_{DS}$. $I_{GS}$-$V_{GS}$ and $I_{DS}$-$V_{GS}$ characteristics under applied $V_{DS}$ = 0.1V, 0.5V, 1V and $V_{GS}$ = ±1V through ±4V are available in Fig. S10. (c,d) Transfer curves of flake MoS$_2$ transferred 250 nm channel devices on 5 nm Al$_{0.72}$Sc$_{0.28}$N with an applied (c) $V_{DS}$ = 0.1V (d) $V_{DS}$ = 1V. $I_{GS}$-$V_{GS}$ and $I_{DS}$-$V_{GS}$ characteristics under applied $V_{DS}$ = 0.1V, 0.5V, 1V and $V_{GS}$ = ±0.5V through ±3V are available in Fig. S11. (e,f) Crossover voltage of flake and MOCVD wet-transferred (e) 500 nm and (f) 100 nm channel (L$_{ch}$) devices on 5 nm Al$_{0.72}$Sc$_{0.28}$N at different applied $V_{DS}$. $I_{GS}$-$V_{GS}$ and $I_{DS}$-$V_{GS}$ characteristics under applied $V_{DS}$ = 0.1 V, 0.5 V, 1 V and $V_{GS}$ = ± 0.5V through ± 4V are available in Fig. S12-S14.

Due to the atomic thickness of 2D transition metal dichalcogenides (TMDC), ambient conditions can modify the electron states of the materials and change their electrical conductivity. For instance, the conductivity of monolayer MoS$_2$ can be enhanced or suppressed by gasses such as O$_2$[31]. The H$_2$O and O$_2$ molecules adsorbed at the surface of the 2D materials then act as interface traps and scattering centers, thereby degrading their electrical properties[22,32]. In addition, the channel-metal interface is also a location of desorption of molecules[32]. We observe this phenomenon in 10 nm thick Al$_{0.68}$Sc$_{0.32}$N devices, where a decrease in switching voltage was observed under vacuum at 1x10$^{-5}$ Torr, both at 300K and 250K (see Fig. S16). We again observe this effect to be independent of gate voltage range or channel length (see Fig. S16(b)), which confirms the behavior is not a contact effect.

**Conclusion & Outlook**

We demonstrate partial ferroelectric switching in ultrathin Al$_{1-x}$Sc$_x$N films (5 nm and 10 nm) integrated with 2D MoS$_2$ channels in FeFET structures. These devices exhibit significantly reduced coercive voltages (<3V), enabling low-voltage operation essential for advanced applications. Our systematic characterization reveals that oxidation disproportionately affects thinner films, while fabrication methods significantly impact device performance, with dry-transferred channels showing superior characteristics. The crossover voltage behavior remains independent of channel length, suggesting it is governed by ferroelectric properties rather than contact effects. Although ferroelectric switching is established, the modest ON/OFF ratio (<5) at zero gate voltage presents a key challenge. Future work should focus on simultaneously improving this ratio while maintaining the low coercive field of these ultrathin films through optimized composition, interface engineering, and fabrication refinements.

**Methods**

The Al/Al$_{1-x}$Sc$_x$N/Al layers were deposited via sputtering onto a *c*-axis-oriented sapphire wafer with an inclination of 0.2° to the M-plane. Sapphire wafers were chosen because of their close lattice match with the template Al, allowing for a highly oriented (111) template layer. The sputtering was completed in-situ without breaking vacuum in an Evatec CLUSTERLINE® 200 II PVD system. The Al is deposited at 150 °C with a target power of 1000 W under 20 sccm. Ar flow, resulting in a process pressure of approximately 1 × 10$^{-3}$ mbar. The Al$_{1-x}$Sc$_x$N layers were deposited via co-sputtering of Al (99.999%) and Sc (99.99%) targets at 88.5 mm from the substrate at 350°C. Various film thicknesses and Sc content are achieved by adjusting the deposition time and target power. The 5 nm Al$_{0.72}$Sc$_{0.28}$N was deposited with Al and Sc target power of 900 W and 555 W under N$_2$ flow 20 sccm. The 10 nm Al$_{0.68}$Sc$_{0.32}$N was deposited with Al and Sc target power of 1000 W and 655 W under N$_2$ flow 30 sccm. The sharpness of the interface between the template Al and Al$_{1-x}$Sc$_x$N under similar growth conditions was shown in previous studies[11]. Finally, the 10 nm AlN was deposited with Al of 900 W under N$_2$ flow 21 sccm. The Al$_{1-x}$Sc$_x$N thicknesses were measured by x-ray reflectometry (XRR). XRR fits were performed with GSAS-II, which uses an optical matrix method as its model. Energy dispersive X-ray spectroscopy (EDS) mapping confirmed the Sc content of the film. Details of the XRR and EDS methods are outlined in a previous study[11].

The choice of bottom template is particularly important for forming highly *c*-axis oriented Al$_{1-x}$Sc$_x$N thin films[33]. Assuming Al$_{0.72}$Sc$_{0.28}$N≈ Al$_{0.68}$Sc$_{0.32}$N≈3.2Å, the Al-Al bond length ($d_{Al}$ = 2.85 Å) is smaller, meaning that the bottom interface shows in-plane compressive stress. Other substrates with closer lattice matches, such as Sc ($d_{Sc}$ = 3.21 Å) remain to be explored.

The top Al layer protects the Al$_{1-x}$Sc$_x$N film from oxidation prior to fabrication of the FeFETs. It is then etched down with Cl$_2$/BCl$_3$ in an Oxford PlasmaPro 100 Cobra inductively coupled plasma (ICP) etcher[34]. The recipe consisted in 1000 W ICP power, 20 W RF power, 8 mTorr chamber pressure, and Cl$_2$/BCl$_3$/Ar flow rate of 5/40/5 sccm, respectively.

Bulk MoS$_2$ purchased from 2D Semiconductors was used for the dry-transfer method. The operation is performed by mechanically exfoliating few-layer MoS$_2$ flakes from the crystal and stamping these onto the Al$_{1-x}$Sc$_x$N surface. MoS$_2$ monolayers were grown by MOCVD (10.60551/znh3-mj13) on c-plane sapphire using Mo(CO)$_6$ and H$_2$S in a H$_2$ carrier gas. Sample growth and characterization are available at the following link: https://data.2dccmip.org/kdhkceelj4Uf.

In the case of chemical wet-transfer of MOCVD grown monolayer MoS$_2$, poly Methyl Methacrylate (PMMA) 950k A4 is spin coated on a piece of MoS$_2$ and dried onto a hot plate at 180°C for < 5 min to prepare for the wet chemical delamination process. The piece is placed in deionized (DI) water at 90° C until air-bubbles begin to form on the surface of the sapphire substrate. Samples are then taken out of the hot water. The delamination of the PMMA-supported MoS$_2$ from its sapphire substrate is performed with the sapphire substrate held manually and slowly dipped inside the KOH solution. Once this step is complete, the PMMA-coated MoS$_2$ monolayer floats at the surface of the KOH solution. Using a clean glass slide, it is rinsed with DI water to remove any residual contamination of KOH. Finally, the PMMA-coated MoS$_2$ monolayer is scooped on the Al/Al$_{1-x}$Sc$_x$N substrate and left to air-dry overnight. To allow for better adhesion of the monolayer to its new substrate, the sample is also heated for 30 min at 75° C on a hot plate. The PMMA film is removed in acetone at 60° C.

To define a consistent channel width in wet-transferred devices, the MoS$_2$ layer is etched in a rectangular shape using a reactive ion etcher (March Jupiter II plasma etcher), set to a power of 100 W, O$_2$ gas flow 450 sccm. and run time 30 s.

Contact metals are defined by electron-beam lithography patterning (Raith EBPG5200+ E-Beam Lithography) followed by thermal evaporation (Kurt J. Lesker Nano-36). The In and Au deposition rates are kept below 0.02 Å/s and 0.5 Å/s respectively. This method minimizes any high-energy damage at the surface of the 2D TMDC[7].

Due to the thinness of our Al$_{1-x}$Sc$_x$N layers, there can be a high leakage between the global back gate and the source/drain contacts. We therefore deposit a lithographically defined insulating Al$_2$O$_3$ layer everywhere around the device channels (see Fig. 1(a)). It consists of a base 2 nm electron-beam deposited Al$_2$O$_3$ layer. An additional ∼38 nm was deposited using a Cambridge Nanotech S200 atomic layer deposition (ALD) system at 100 °C. The Al$_2$O$_3$ is also patterned using a Raith EBPG5200+ E-Beam Lithography writer. This fabrication flow was not modified when repeated on an Al/AlN/Al substrate for the reference device presented in Fig. 2(d).

$I_{DS} - V_{GS}$ and leakage $I_{GS} - V_{GS}$ measurements are carried out simultaneously using a Keithley 4200A-SCS Parameter Analyzer. We know that if ferroelectric film dipoles rotate slower than the observation time during transfer measurements, no hysteresis will occur. Varying voltage steps would also change the charge injection during transfer measurements, leading to different tunneling depths. For this reason, we maintained $V_{GS}$ sweeps in incremental steps of 0.01 V during measurements. High vacuum measurements at both 300K and 250K were performed with the same analyzer in a cryogenic probe station (Lakeshore CRX-4K).

**Conflicts of interest**

The authors declare no competing financial interests.

**Acknowledgements**

C. L., Y. Z. H.C., R. H. O., and D. J. acknowledge support for this work from Intel Corporation under the SRS program. The authors also gratefully acknowledge the help and support provided by Shriram Shivaraman from Intel. S. S. acknowledges support for this work by Basic Science Research Program through the National Research Foundation of Korea (NRF) funded by the Ministry of Education (Grant Number 2021R1A6A3A14038492). The MoS$_2$ monolayer samples were provided by the 2D Crystal Consortium–Materials Innovation Platform (2DCC-MIP) facility at the Pennsylvania State University which is funded by NSF under cooperative agreement DMR-2039351. A portion of work was carried out in part at the Singh Center for Nanotechnology, which is supported by the NSF National Nanotechnology Coordinated Infrastructure Program under Grant NNCI-2025608.

# Supplementary Information:
# Low-voltage Ferroelectric Field-Effect Transistors with Ultrathin Aluminum Scandium Nitride and 2D channels

Chloe Leblanc[1], Hyunmin Cho[1], Yinuo Zhang[1], Seunguk Song[1,2,3], Zachary Anderson[1], Yunfei He[1], Chen Chen[4], Joan Redwing[4], Roy H. Olsson, III[1], Deep Jariwala[1, *]

[1]Department of Electrical and Systems Engineering, University of Pennsylvania, Philadelphia, Pennsylvania, USA
[2]Department of Energy Science, Sungkyunkwan University (SKKU), Suwon, 16419, Republic of Korea
[3]Center for 2D Quantum Heterostructures (2DQH), Institute for Basic Science (IBS), Sungkyunkwan University (SKKU), Suwon, 16419, Republic of Korea
[4]2D Crystal Consortium–Materials Innovation Platform, Materials Research Institute, Pennsylvania State University, University Park, Pennsylvania, USA

[*]Author to whom correspondence should be addressed: dmj@seas.upenn.edu.

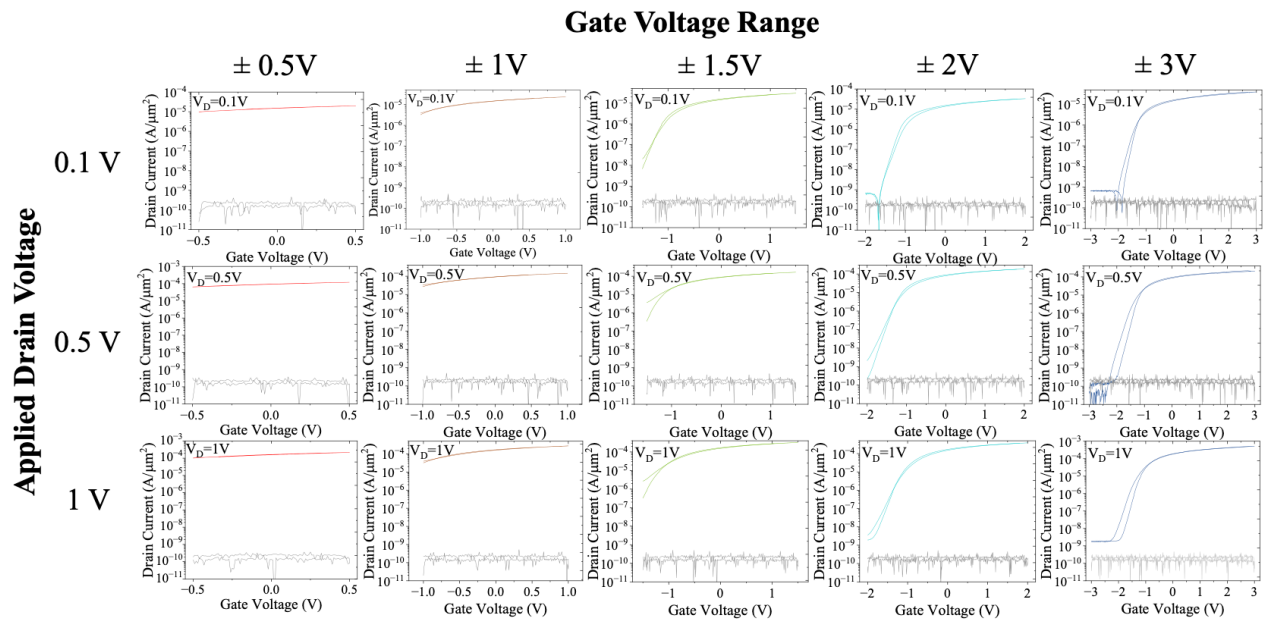

**Figure S1** $I_{GS}$-$V_{GS}$ and $I_{DS}$-$V_{GS}$ characteristics of the dry-transferred 50 nm channel flake MoS$_2$ device on 5 nm Al$_{0.72}$Sc$_{0.28}$N presented in Fig. 1 on logarithmic scale. Leakage $I_{GS}$ is plotted in gray. Voltage increment is 0.01V step size, and currents are normalized by channel area.

Supporting Information

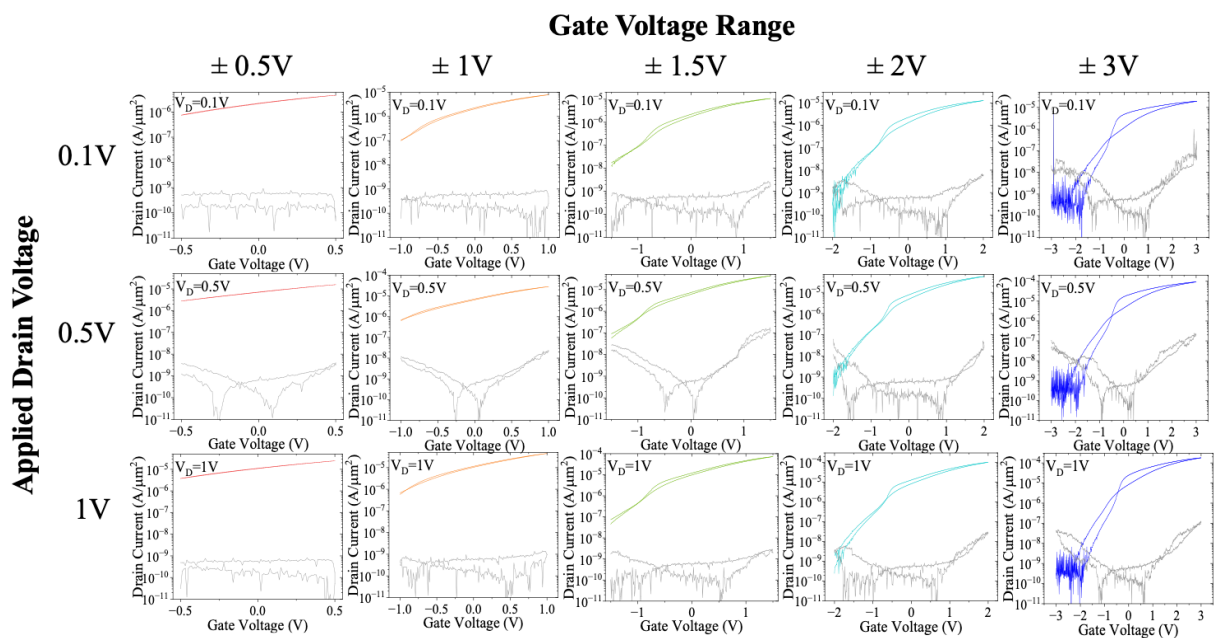

**Figure S2** $I_{GS}$-$V_{GS}$ and $I_{DS}$-$V_{GS}$ characteristics of the dry-transferred 50 nm channel flake MoS$_2$ device on 10 nm Al$_{0.68}$Sc$_{0.32}$N presented in Fig. 1 on logarithmic scale. Gray curves are the leakage gate current $I_{GS}$. Voltage increment is 0.01V step size, and currents are normalized by channel area.

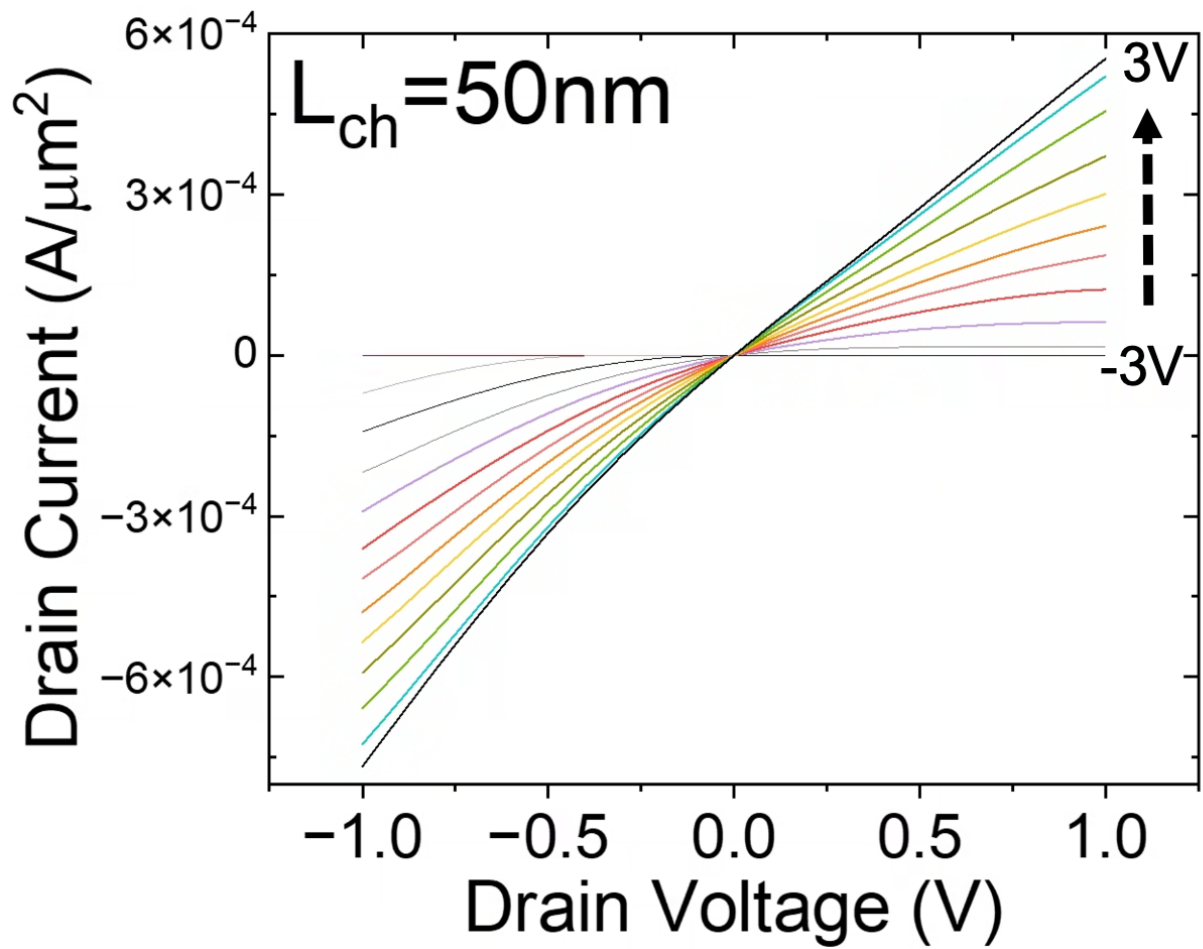

**Figure S3** Output characteristics of the 50 nm channel ($L_{ch}$) device on 5 nm $Al_{0.72}Sc_{0.28}N$ presented in Fig. 1. The Ohmic contacts allow for a high injection of electrons from the In/Au to the $MoS_2$ channel (see Methods). Currents are normalized by channel area

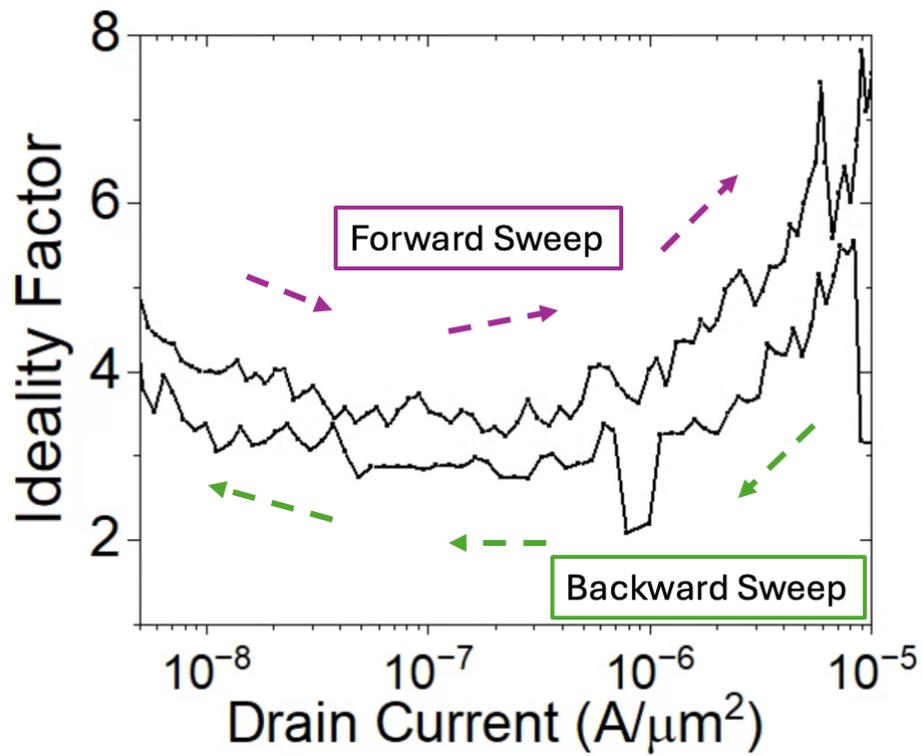

**Figure S4** Ideality factor vs. $I_{DS}$ characteristics during forward and reverse sweeps of the 5 nm $Al_{0.72}Sc_{0.28}N$ device in Fig. 2a.

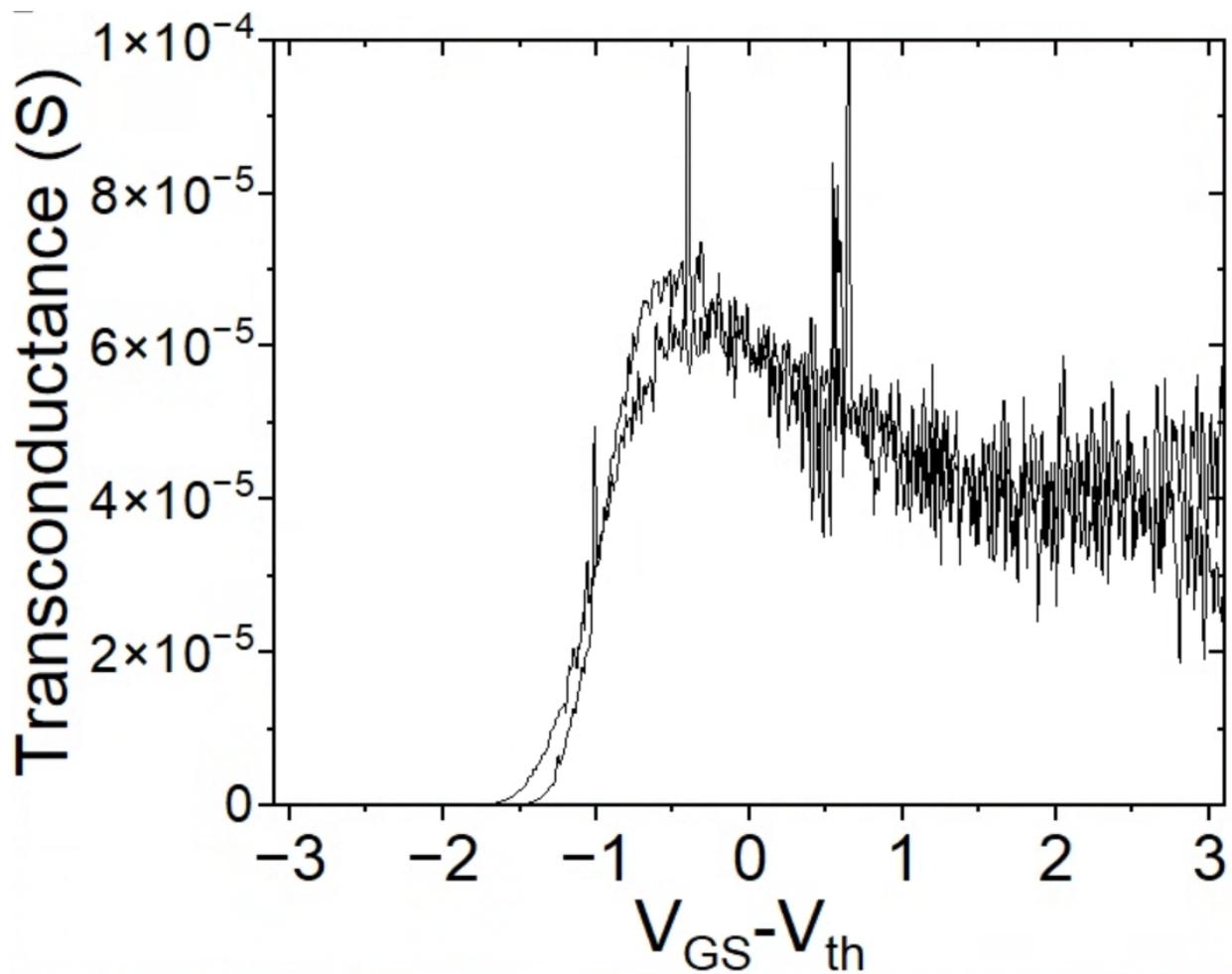

**Figure S5** Transconductance vs. $V_{GS}$ -$V_{th}$ characteristics during forward and reverse sweeps of the 5 nm $Al_{0.72}Sc_{0.28}N$ device in Fig. 2(a). Removing the aberrations, we record at maximum transconductance value $g_{m,max}$ = 7.36 x $10^{-5}$ S at gate voltage $V_{GS}$ - $V_{th}$ = -0.49 + 0.17 = -0.32 V during the backward measurement sweep.

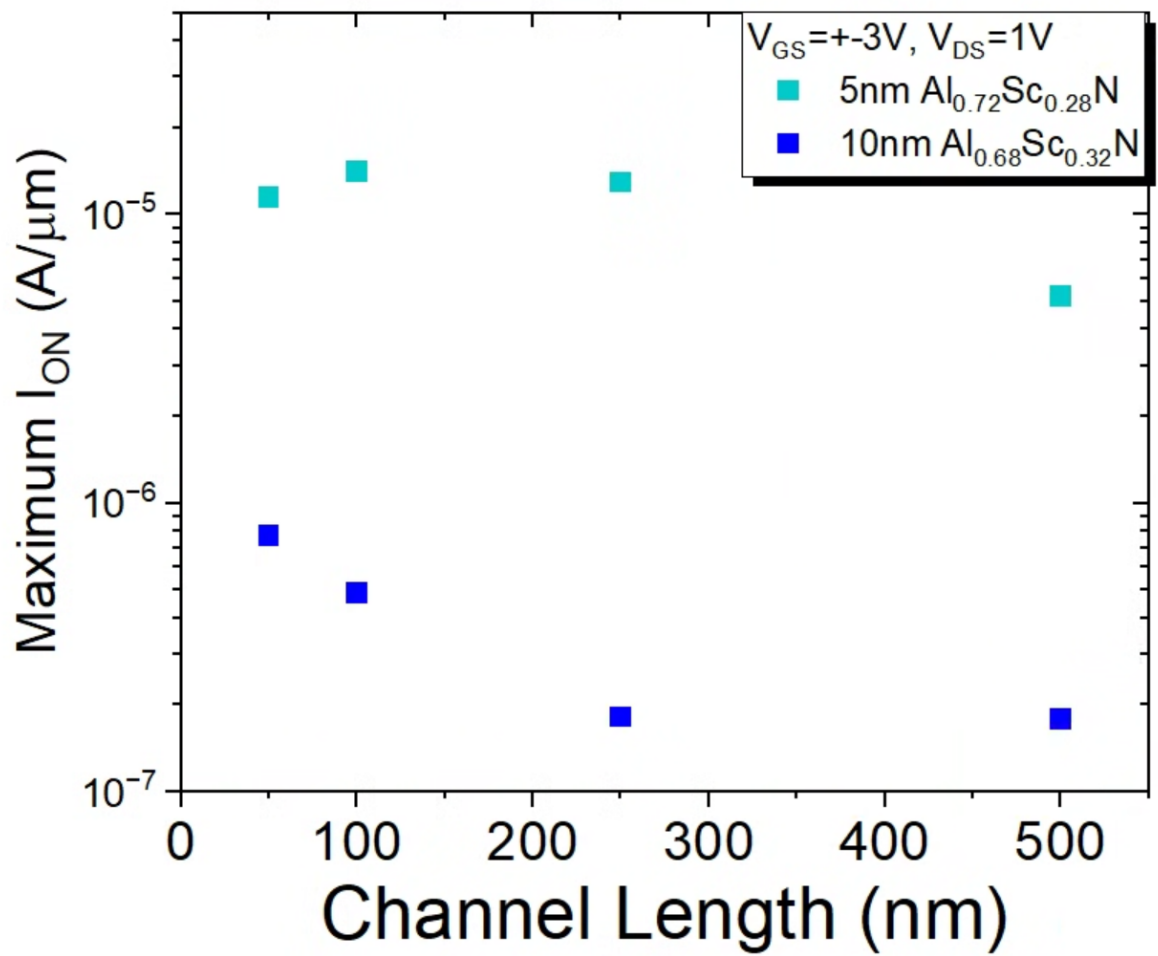

**Figure S6** Maximum ON current normalized by channel width in wet-transferred 5 nm thick $Al_{0.72}Sc_{0.28}N$ and 10 nm thick $Al_{0.68}Sc_{0.32}N$ FeFETs. $V_{GS}$ range is ±3V and the applied $V_{DS}$ = 1V.

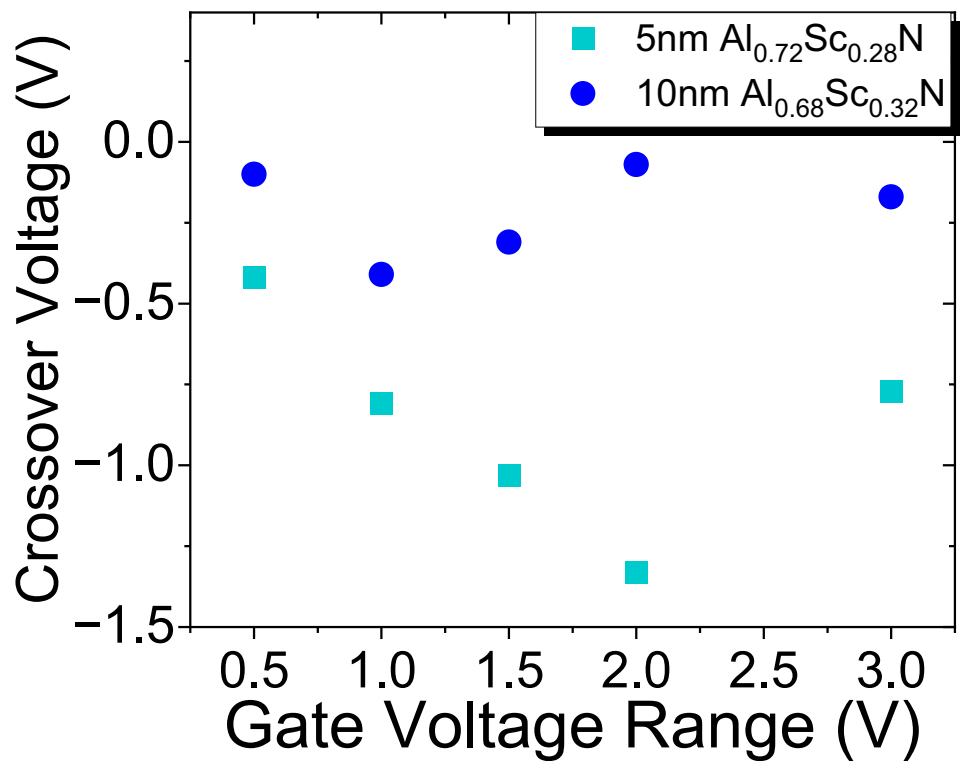

**Figure S7** Crossover voltage of the 5 nm thick $Al_{0.72}Sc_{0.28}N$ and 10 nm thick $Al_{0.68}Sc_{0.32}N$ FeFETs presented in Fig. 2. $V_{GS}$ ranges ±0.5V through ±3V and applied $V_{DS}$ = 1V. Voltage increment is 0.01V step size.

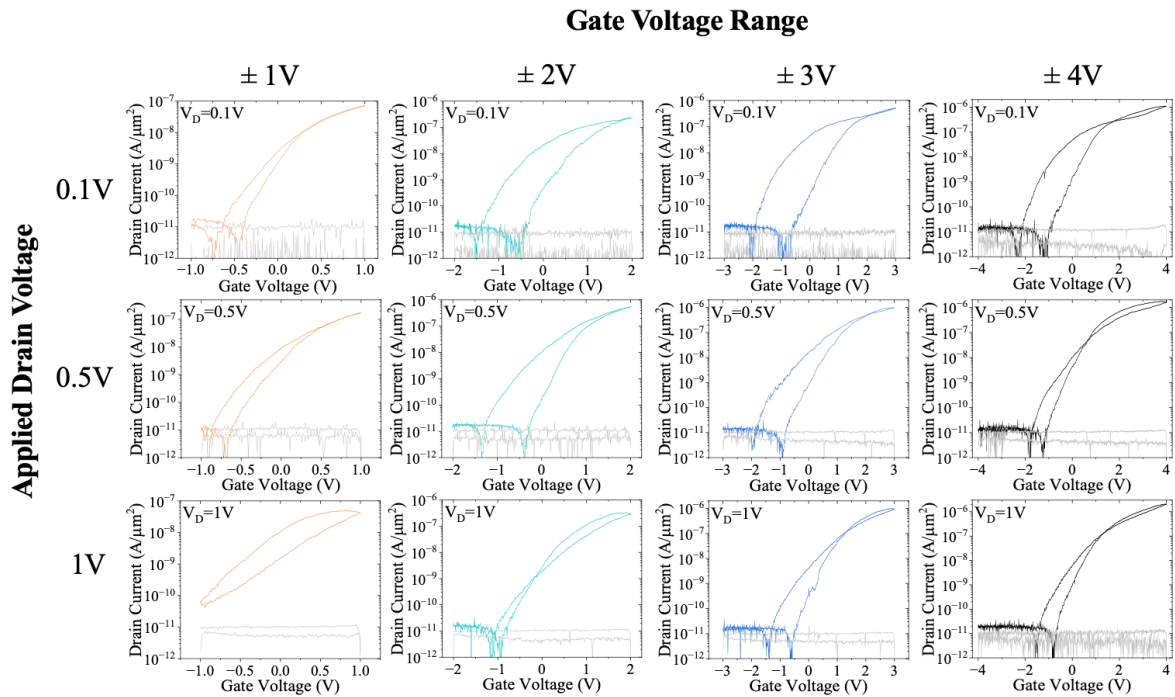

**Figure S8** $I_{GS}$-$V_{GS}$ and $I_{DS}$-$V_{GS}$ characteristics of a wet-transferred MOCVD MoS$_2$ 500 nm channel device on 10 nm Al$_{0.68}$Sc$_{0.32}$N on logarithmic scale. Gray curves are the leakage gate current $I_{GS}$. Voltage increment is 0.01V step size, and currents are normalized by channel area.

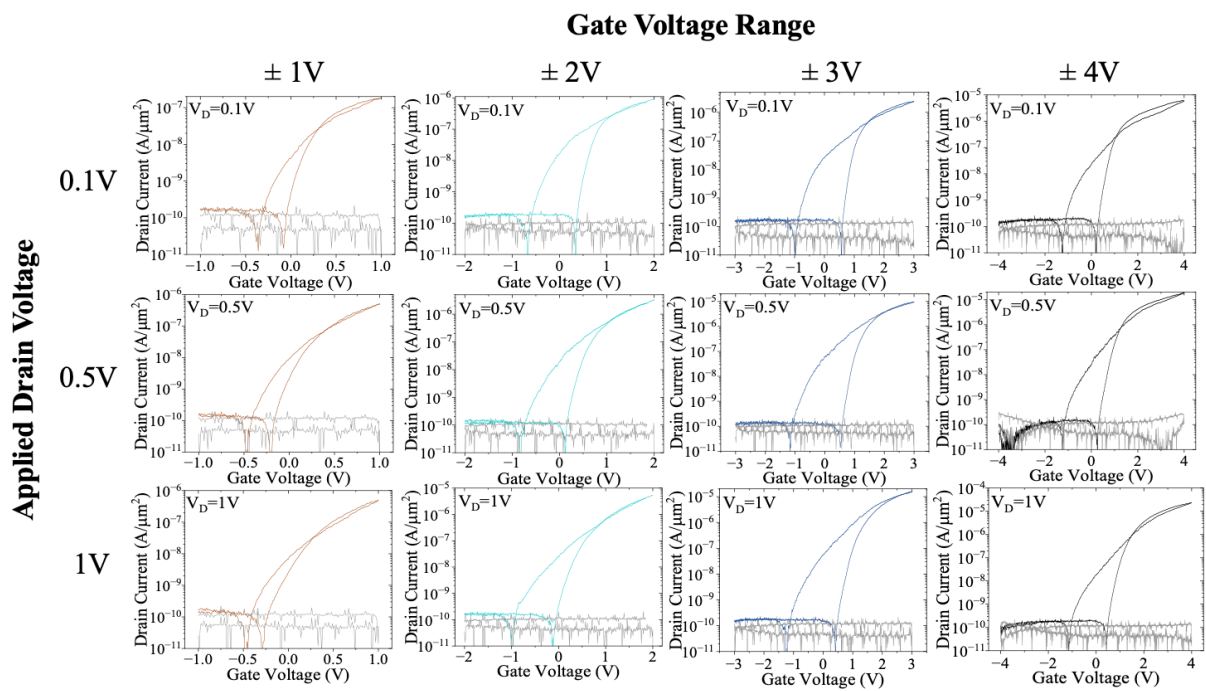

**Figure S9** $I_{GS}$-$V_{GS}$ and $I_{DS}$-$V_{GS}$ characteristics of a wet-transferred MOCVD MoS$_2$ 50 nm channel device on 10 nm Al$_{0.68}$Sc$_{0.32}$N on logarithmic scale. Gray curves are the leakage gate current $I_{GS}$. The crossover voltages are all shifted in comparison to the dry-transferred device presented in Fig. S2. Voltage increment is 0.01V step size, and currents are normalized by channel area.

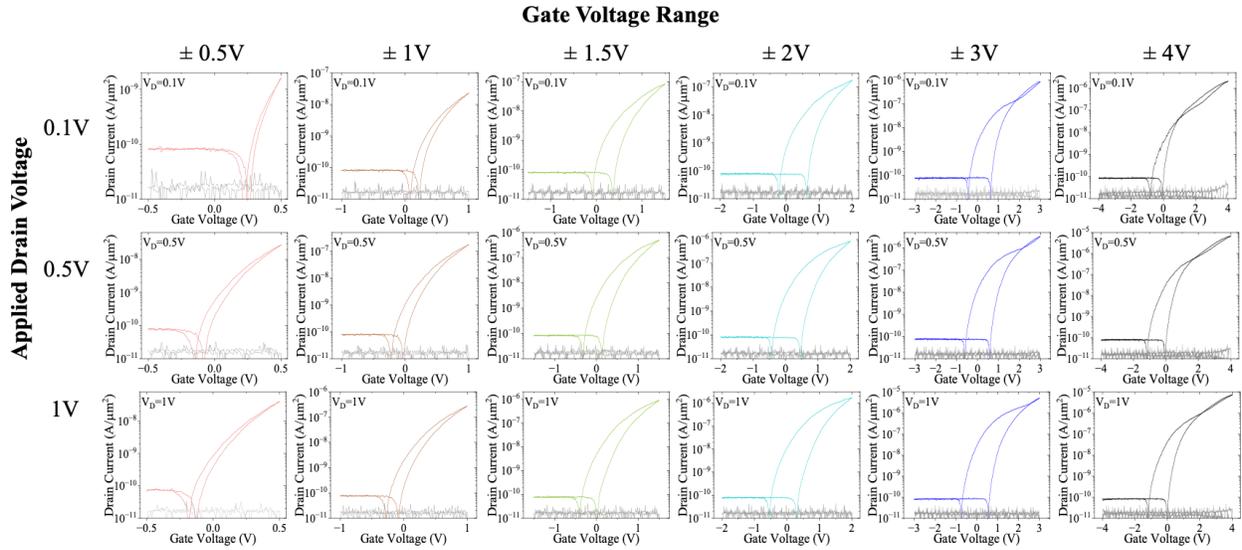

**Figure S10** $I_{GS}$-$V_{GS}$ and $I_{DS}$-$V_{GS}$ characteristics of the wet-transferred MOCVD MoS$_2$ 250 nm channel device on 5 nm Al$_{0.72}$Sc$_{0.28}$N on logarithmic scale. Gray curves are the leakage gate current $I_{GS}$. Voltage increment is 0.01V step size, and currents are normalized by channel area.

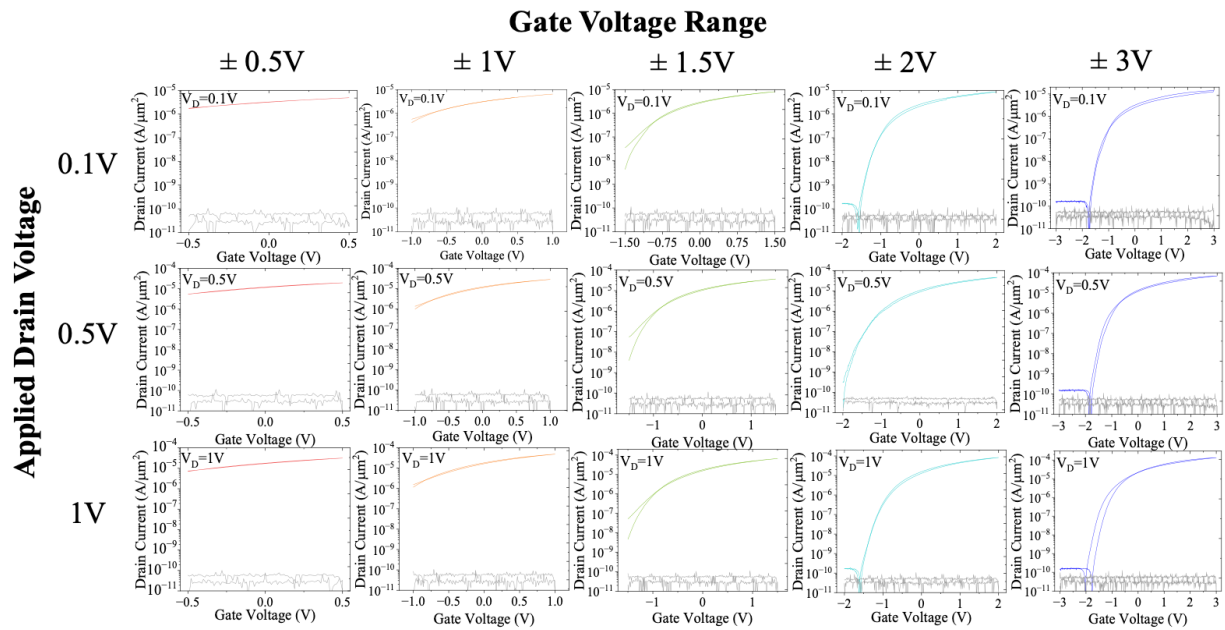

**Figure S11** $I_{GS}$-$V_{GS}$ and $I_{DS}$-$V_{GS}$ characteristics of the dry-transferred flake 250 nm channel device on 5 nm $Al_{0.72}Sc_{0.28}N$ presented in Fig. 3(d,e) on logarithmic scale. Gray curves are the leakage gate current $I_{GS}$. Voltage increment is 0.01V step size, and currents are normalized by channel area.

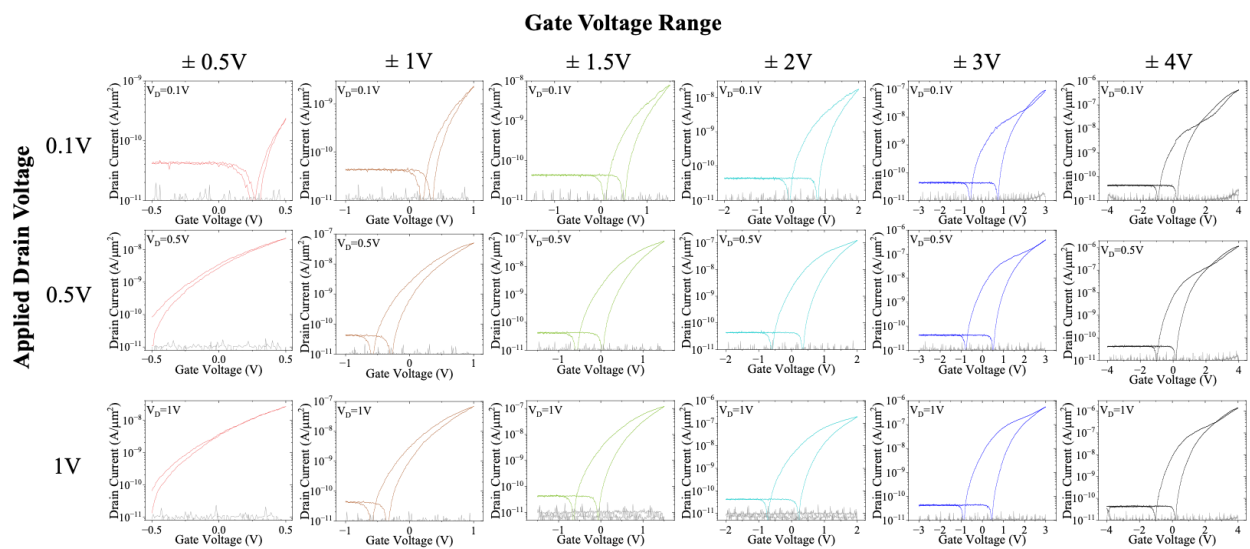

**Figure S12** $I_{GS}$-$V_{GS}$ and $I_{DS}$-$V_{GS}$ characteristics of the wet-transferred MOCVD MoS$_2$ 500 nm channel device on 5 nm Al$_{0.72}$Sc$_{0.28}$N on logarithmic scale. Gray curves are the leakage gate current $I_{GS}$. Voltage increment is 0.01V step size, and currents are normalized by channel area.

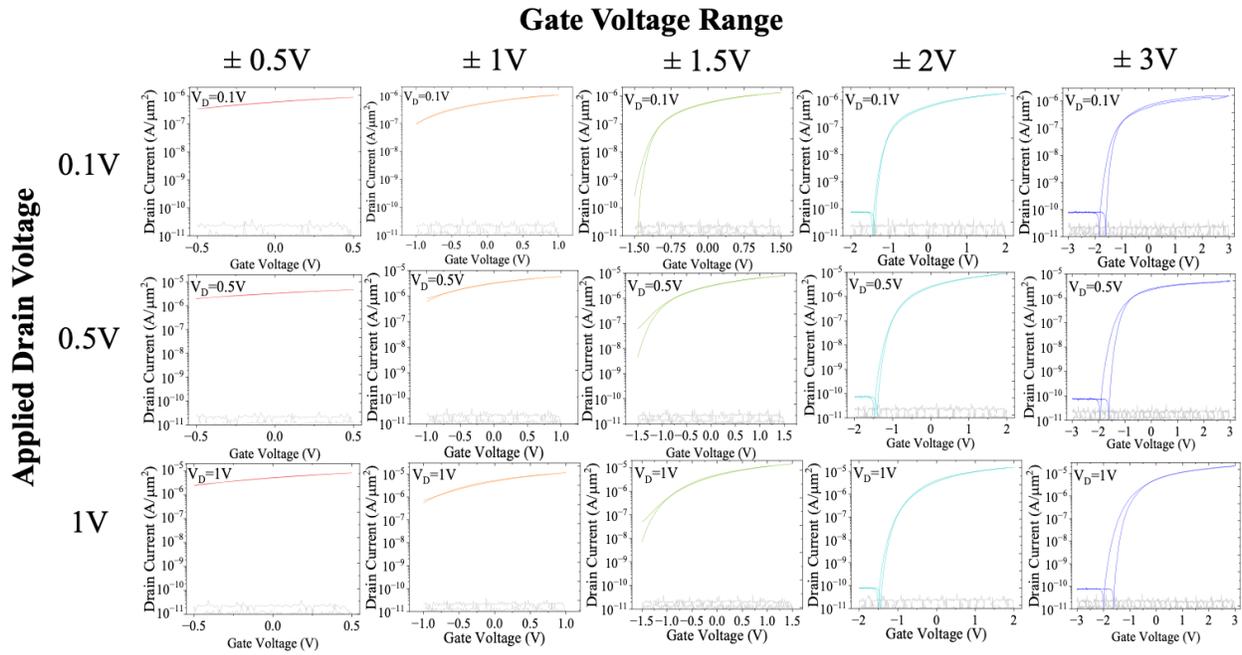

**Figure S13** $I_{GS}$-$V_{GS}$ and $I_{DS}$-$V_{GS}$ characteristics of the dry-transferred flake MoS$_2$ 500 nm channel device on 5 nm Al$_{0.72}$Sc$_{0.28}$N on logarithmic scale. Gray curves are the leakage gate current $I_{GS}$. Voltage increment is 0.01V step size, and currents are normalized by channel area.

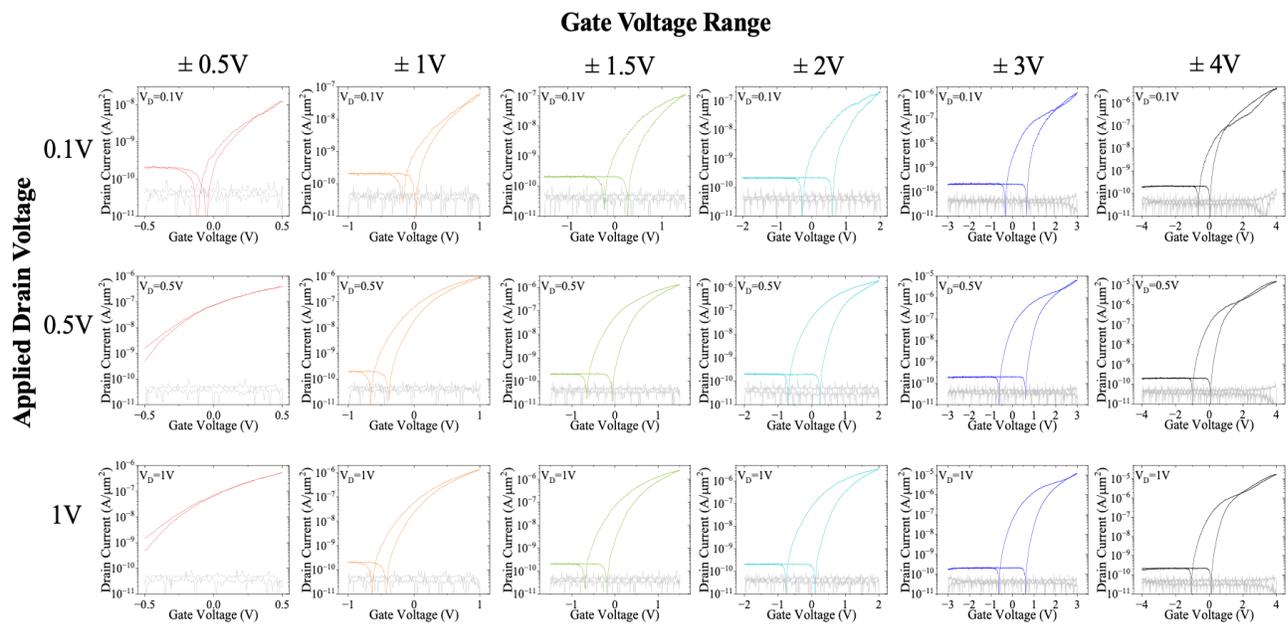

**Figure S14** $I_{GS}$-$V_{GS}$ and $I_{DS}$-$V_{GS}$ characteristics of a wet-transferred MOCVD MoS$_2$ 100 nm channel device on 5 nm Al$_{0.72}$Sc$_{0.28}$N on logarithmic scale. Gray curves are the leakage gate current $I_{GS}$. Voltage increment is 0.01V step size, and currents are normalized by channel area.

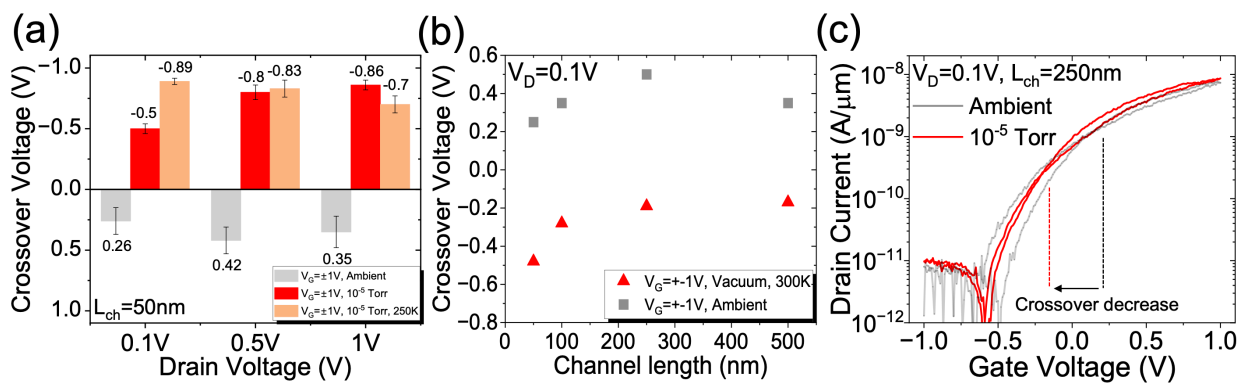

**Figure S16** (a) Crossover voltage of wet-transferred 50 nm channel ($L_{ch}$) devices on 10 nm $Al_{0.68}Sc_{0.32}N$ at different applied $V_{DS}$ and ambient conditions. (b) Crossover voltage of wet-transferred devices on 10 nm $Al_{0.68}Sc_{0.32}N$ with applied $V_{DS}$ = 0.1V and $V_{GS}$ = ±3V at different channel lengths in ambient and vacuum conditions. (c) Transfer curves of wet-transferred 250 nm channel devices on 10 nm $Al_{0.68}Sc_{0.32}N$ under ambient conditions and vacuum. Voltage increment is 0.01V step size.

Supporting Information